\documentstyle[preprint,eqsecnum,aps]{revtex}
\begin{document}
\draft
\everymath={\displaystyle}
\preprint{IPNO/TH 95-38}
\title{Strongly Coupled Positronium in a Chiral Phase}
\author{M. Bawin and J. Cugnon}
\address{Universit\'{e} de Li\`{e}ge, Institut de Physique B5, Sart
Tilman, B-4000 Li\`{e}ge 1, Belgium}
\author{H. Sazdjian} \address{Division de
Physique Th\'{e}orique \thanks{Unit\'e de Recherche des Universit\'es Paris 11
et Paris 6 associ\'ee au CNRS.}, Institut de Physique Nucl\'{e}aire,
Universit\'{e} Paris XI,\\ F-91406 Orsay Cedex, France\\ \vspace{11 cm} }

\maketitle
\pagebreak
\begin{abstract}
Strongly coupled positronium, considered in its pseudoscalar sector,
is studied in the framework of relativistic quantum constraint dynamics.
Case's method of self-adjoint extension of singular potentials, which
avoids explicit introduction of regularization cut-offs, is adopted.
It is found that, as the coupling constant $\alpha $ increases, the bound
state spectrum undergoes an abrupt change at the critical value
$\alpha =\alpha _c =1/2$. For $\alpha >\alpha _c$, the mass spectrum
displays, in addition to the existing states for $\alpha <\alpha _c$,
a new set of an infinite number of bound states concentrated in a
narrow band starting at mass $W=0$; all the states have indefinitely
oscillating wave functions near the origin. In the limit $\alpha \rightarrow
\alpha _c$ from above, the oscillations disappear and the narrow band of
low-lying states shrinks to a single massless state with a mass gap with
the rest of the spectrum. This state has the required properties to
represent a Goldstone boson and to signal spontaneous breakdown of chiral
symmetry. It is suggested that the critical coupling constant $\alpha _c$
be viewed as a possible candidate for an ultra-violet stable fixed point
of QED, with a distinction between two phases, joined to each other by a
first-order chiral phase transition.
\end{abstract}
\pacs{PACS numbers : 03.65.Pm, 11.10.St, 12.20.$-$m, 11.30.Rd.}

\narrowtext

\section{INTRODUCTION}

The question of a possible existence of an ultra-violet stable fixed point
in QED was investigated long ago by Gell$-$Mann and Low \cite{gml} and
developed later by several authors \cite{bjw,ad,adb}. If such a point were
to exist, then the electron mass would be entirely
dynamical in origin \cite{bjw,ad,adb}, with a possible spontaneous breakdown of
chiral symmetry \cite{p}.
Although perturbation theory calculations do not seem to point to the
existence of such a solution, quenched lattice QED calculations
displayed the existence of a phase transition at the critical value
$\alpha _c \sim 0.3$ of the coupling constant $\alpha $, with the
occurrence of a spontaneous breakdown of chiral symmetry \cite{b}. These
observations were also confirmed with unquenched lattice calculations \cite{g},
with $\alpha _c \sim 0.4$, but a vanishing of the Callan-Symanzik
function $\beta $ was not found there and the question of the validity
of QED as a nontrivial consistent theory in the continuum limit was
raised.\par
On the other hand, it is possible, in the continuum theory, to analyze
a partial, but simpler, aspect of the phase transition problem, namely
that of the bound state of strongly coupled positronium. If the
ground state of the corresponding spectrum, for some value of
$\alpha $, were massless, then this would be the signal of a possible
spontaneous breakdown of chiral symmetry and the prelude of
a phase transition in QED. This result would not be, however, sufficient by
itself to ensure the vanishing of the Callan-Symanzik $\beta$-function,
which should be shown by independent calculations; only in this case
could the consistency of the whole procedure be guaranteed.\par
We studied, in previous work \cite{bc,bcs}, the problem of strongly
coupled positronium in the framework of relativistic quantum constraint
dynamics (RQCD) \cite{cvawb,sms}. This framework provides a manifestly
covariant three-dimensional description of the internal motion of
two-body systems and can be shown to be equivalent to
a three-dimensional reduction of the Bethe$-$Salpeter equation \cite{s}.
When the approximation of local potentials is made, the corresponding
wave equations can be analyzed rather easily and in many cases analytic
solutions can be obtained.
It was found that the Todorov form of the electromagnetic two-body
potential, first introduced in the quasipotential approach \cite{t},
leads to the existence of a critical value of the coupling constant,
with $\alpha _c =1/2$. For $\alpha > \alpha _c $, the potential becomes
too singular and needs some regularization, provided by a
cut-off radius $r_{0}$ in the Coulomb interaction.
In the regularized theory  the bound state mass spectrum displays, for $\alpha
\gtrsim 1/2$, a rapid fall of the ground state mass to values close
to zero, thus indicating a drastic modification of the qualitative features
of the bound system and presumably of the theory itself.
The above procedure requires, however, a numerical treatment of the
equations and makes it difficult to ascertain the existence of solutions for
vanishing values of $r_{0}$.\par
The presence of a cut-off radius in the Coulomb
interaction needs, in general, an adequate interpretation.
As long as one studies the behavior of charged particles in
supercritical Coulomb fields of heavy nuclei, the finite size of the latter
naturally regularizes the Coulomb interaction at short distances.
If, on the other hand, it is the strong interaction of pointlike particles
that is considered, as in the case of positronium in QED with strong
Coulomb coupling, then the meaning of $r_0$ remains unclear.\par
A similar problem also occurs with the Bethe-Salpeter equation in the
ladder approximation. It has been shown \cite{mnfk} that, when the coupling
constant $\alpha $ is larger than a critical value $\alpha _c$ ($\sim
\pi /3$ in the Landau gauge and $\sim \pi /4$ in the Feynman gauge),
the theory undergoes spontaneous breakdown of chiral symmetry.
However, for $\alpha \gtrsim \alpha _c$, the treatment and resolution of the
equation necessitate the use of an ultra-violet cut-off $\Lambda $.
While the introduction of the latter can naturally be justified in QCD
as being an approximate way of parametrizing the asymptotic freedom of the
theory \cite{h}, it has not received a simple interpretation in QED. In this
respect, Miransky {\it et al.} \cite{fgmmmf} suggested that, for $\alpha >
\alpha _c$, QED undergoes an additional charge renormalization that
absorbs the infinities of the pointlike limit; the renormalized charge
remains equal to $\alpha _c$ and then might be identified with the
ultra-violet fixed point of QED.
It was also pointed out in this connection \cite{llb,kdk} that,
for $\alpha = \alpha _c$, because of the new
renormalization of $\alpha $, the fermion composite operator $\overline
\psi \psi$ acquires the dimension 2 instead of 3, and thus allows
for the presence of renormalizable four-fermion interactions.\par
In quantum mechanics, there exists an alternative
method for dealing with singular interactions of pointlike particles,
without the need of introducing cut-offs: this is the self-adjoint
extension method, discussed a long time ago for singular potentials of the
type $1/r^n$ ($n \geq 2$) by Case \cite{c}.
In his classic paper, Case showed that all self-adjoint
extensions of the Klein-Gordon-Coulomb (or Dirac-Coulomb) problem can be
parametrized by a single constant $B$ when the interaction becomes singular.
\par
While Case's method is only of academic interest for problems concerning
supercritical Coulomb fields of heavy nuclei, the size of the latter
providing a natural short distance cut-off, it reveals its full power
in the present problem of strong Coulomb interaction of pointlike particles.
The self-adjoint extension parameter $B$ can be interpreted as
parametrizing the short distance behavior of the interaction and its
choice amounts to fixing the energy of one of the bound states (the values
of the masses of the constituent particles  of the bound states and
of the coupling
constant $\alpha $ being already fixed) and calculating the other bound state
energies with respect to this one, without making explicit cut-offs appear.
\par
In the case of $\delta $-function interactions in two and three space
dimensions, it was shown \cite{j} that the self-adjoint extension method
provides the renormalized version of the theory, when the cut-off of the
regularized theory is removed and a corresponding renormalization of the
coupling constant is performed. It is then natural to expect from the
same method of approach, applied now to the $1/r^2$ singularity, to also
provide the finite renormalized version of the theory, provided one of the
bound state energies is fixed.\par
Motivated by these results, we have investigated with Case's method of
self-adjoint extension the problem of strong Coulomb coupling in
positronium-like systems. The relativistic wave equations of constraint
theory lead for the relative motion in $^1S_0$ states to a final
three-dimensional equation which is very similar in form to the
Klein-Gordon (KG) equation \cite{bcs}, and therefore Case's method can
be readily applied to it.\par
Our main results are the following. We find that the system undergoes a
first-order chiral
phase transition at the critical value $\alpha = \alpha _c =1/2$. While
the ground state mass for $\alpha < 1/2$ can be continued to the domain
$\alpha >1/2$ and remains different from zero, a new set of an infinite
number of states, concentrated in a finite domain of mass with
accumulation at the value zero, appears, the zero mass state
representing the new ground state of the system. This result occurs for
any fixed value of Case's constant $B$.
All the states have indefinitely oscillating wave functions near the
origin. While tachyonic solutions formally exist, they are ruled out
from the spectrum by the self-adjointness condition, and therefore the
zero mass state remains the physical ground state of the spectrum.\par
In the limit $\alpha \rightarrow \alpha _c$ from above, the
short-distance oscillations disappear from the wave functions, and
the states accumulated around the zero mass solution shrink to a single
massless state with a definite mass gap with the rest of the spectrum.
It turns out that the latter state has the required properties to
represent a Goldstone boson and hence to signal a spontaneous
breakdown of chiral symmetry. The fact that for $\alpha >\alpha _c$ a
sensible theory, with finite and nonvanishing couplings to the
observable currents, can be defined only for $\alpha =\alpha _c +0$
strongly suggests the identification of $\alpha _c$ with an
ultra-violet stable fixed point of QED, with the distinction between two
phases, governed by $\alpha _c -0$ and $\alpha _c +0$, respectively,
and joined to each other by a first-order chiral phase transition.
\par
The paper is organized as follows.  Section II discusses Case's
method for Klein-Gordon  particles in an external Coulomb field and is
included to make the paper self-contained.  Section III is devoted to the
study, in the framework of RQCD, of the strongly coupled positronium
spectrum in its pseudoscalar sector. Section IV deals with
the question of an eventual appearance of tachyonic states in the
spectrum of states. In Sec. V, the limit $\alpha \rightarrow \alpha _c
+0$ is considered and the presence of a Goldstone boson established. Summary
and discussion of results follow in Sec. VI. In the appendix, some results
of RQCD are summarized.\par

\section{CASE'S METHOD FOR THE KLEIN-GORDON EQUATION}

For the sake of completeness and for easy comparison with RQCD, we discuss here
Case's method\cite{c} for a Klein-Gordon particle of charge $e$, mass $M$
and energy
$E$ in a Coulomb potential $V(r) = - {\alpha Z}/{r}$.  We have $\alpha$ =
1/137 and $\hbar = c = 1$ in our system of units.  The radial KG equation for
$s$-states is :
\begin{equation}
u''\ +\ \left[{\ {E}^{\ 2}\ -\ {M}^{\ 2}\ +\ {\frac{2\ E\ \alpha \rm \ Z}{r}}\
+\ {\frac{{\alpha }^{\ 2}\ {Z}^{\ 2}}{{r}^{\ 2}}}\ }\right]\ u\ =\ 0\ \ .
\end{equation}
{}From Eq. (2.1) one derives the orthogonality condition
for two solutions $u_{1}$ and $u_{2}$ of energy $E_{1}$ and
$E_{2}$ ($E_1 \neq E_2$) :
\begin{eqnarray}
{\left({{u}_{1},{u}_{2}}\right)}_{KG}\ &=&\ {\frac{1}{2}}\
\int_{0}^{\infty }\left({{E}_{1}+{E}_{2}+{\frac{2\alpha Z}{r}}}\right)\
{u}_{1}{u}_{2}\ dr\nonumber \\
&=&\  {\frac{1}{2}}\
{{\left({{u}_{2}{'}{u}_{1}-{u}_{1}{'}{u}_{2}}\right)}\bigg \vert }_{0}^
{\infty}\ =\ 0\ \ \ .
\end{eqnarray}
Note further that the KG norm is given by
\begin{equation}
{\left({u , u}\right)}_{KG}\ =\ \int_{0}^{\infty }\left({E\ +\ {\frac{\alpha
Z}{r}}}\right)\ {u}^{2}\ dr\ \ .
\end{equation}
\par
For $\alpha Z>1/2$, the square integrable solutions of Eq. (2.1)
(vanishing at $r = \infty$) are given by
\begin{equation}
u(E,r)\ =\ c {W}_{ k,\mu} (\rho ) \ ,
\end{equation}
where $c$ is a normalization constant, $W_{k,\mu}$ is the Whittaker function
\cite{bc} and where
\begin{equation}
\rho \ =\ 2 ( {M}^{ 2} - {E}^{ 2} )^{\frac {1}{2}} r\ ,
\end{equation}
\begin{equation}
k\ =\ {\frac{E \alpha  Z}{( M^2 - E^2)^{\frac{1}{2}}}}\ ,
\end{equation}
\begin{equation}
\mu\ =\ i\ \lambda \  ,\ \ \ \lambda \ =\ ( {\alpha }^{ 2}
{Z}^{ 2} - {\frac{1}{4}})^{\frac{1}{2}}\  .
\end{equation}
\par
Using the formula \cite{as1} :
\begin{equation}
\lim _{\rho \rightarrow 0} {W}_{k,\mu } ({\rho })\ =\
{\frac{\Gamma \left({-2\mu
}\right)}{\Gamma \left({{\frac{1}{2}} - \mu  - k}\right)}}
\ {\rho}^{{\frac{1}{2}} + \mu }\ +\ {\frac{\Gamma \left({2\mu }
\right)}{\Gamma \left({{\frac{1}{2}} + \mu  - k}\right)}}
\ {\rho }^{{\frac{1}{2}} - \mu }\  ,
\end{equation}
one finds :
\begin{equation}
\lim _{\rho \rightarrow 0} u\ \sim \ {\rho }^{\frac{1}{2}}
\cos \left({\beta  + \lambda  \ln \rho }\right)\ ,
\end{equation}
with
\begin{equation}
\beta \ =\ \arg\ {\frac{\Gamma \left({ - 2 i \lambda }\right)}
{\Gamma \left({ {\frac{1}{2}} - i \lambda  - k}\right)}}\ .
\end{equation}
It is easy to verify that the general behavior of the
solution of Eq. (2.1) close to $r=0$ is given by \cite{c} :
\begin{equation}
\lim _{r\rightarrow 0} u \sim { {r}^{{\frac{1}{2}}} \cos \left(\lambda
\ln (Mr)  +  B\right)}\ ,
\end{equation}
where $B$ is an arbitrary constant.
Comparison of Eqs. (2.9) and (2.11) yields the relation :
\begin{eqnarray}
\arg \Gamma  \left({ 1 - 2 i \lambda  }\right)\
&+&\ {\frac{\pi }{2}}\ -\ \arg \Gamma  \left({ {\frac{1}{2}} - i\
\lambda  - k }\right)\ +\ \lambda  \ln 2\nonumber \\
&+&\ \lambda  \ln  {\frac{{\left({ {M}^{ 2} -
 {E}^{ 2} }\right)\ }^{{\frac{1}{2}}}}{M}}\ =\ B\ +\ n\pi \ ,
\end{eqnarray}
where $n$ is an arbitrary integer.  Furthermore, it can be shown
\cite{c} that keeping the same value of
$B$ for all states guarantees the orthogonality condition (2.2).
Therefore, choosing the
value of $B$ provides a self-adjoint extension of the KG equation and allows us
to obtain the corresponding spectrum by solving Eq. (2.12) for $E$. Notice
that this equation is invariant under adding to $B$ any multiple of $\pi$ and,
therefore, it is sufficient to consider the values of
$B$ in the interval
$[0, \pi]$.\par
It is of interest
to investigate the small $\lambda$ behaviour of (2.12).  Using the
formula\cite{as2} :
\begin{equation}
\arg \Gamma  \left({\ x\ +\ iy\ }\right)\ =\ y\ \psi \ (x)\ +
\sum_{n\ =\ 0}^{\infty } \left({\ {\frac{y}{x\ +\ n}}\ -
\ \arctan\
{\frac{y}{x\ +\ n}}\ }\right)\  ,
\end{equation}
where
\begin{equation}
\psi  \left({ x }\right)\ =\ {\frac{\Gamma ' \left({ x\
}\right)}{\Gamma  \left({ x }\right)}}\  ,
\end{equation}
one finds from Eq. (2.12) :
\begin{equation}
-2\lambda \psi ({1})\ +\ {\frac{\pi }{2}}\ +\ \lambda
\psi  ({{\frac{1}{2}}-k})\ +\ \lambda  \ln 2 \sqrt
{1-{\frac{{E}^{2}}{{M}^{2}}}}\ +\ {\cal O} \left({{\lambda }^{2}}\right)\ =\
B\ \ .
\end{equation}
(Without loss of generality, we can drop here the $n\pi$ term for small
values of $\lambda$.) For this equation to be satisfied
(for $B \neq \frac{\pi}{2})$, as
$\lambda \rightarrow 0$, it is necessary that
\begin{equation}
\lim _{\lambda \rightarrow 0} \lambda \psi ({{\frac{1}{2}} - k})\
=\ B - {\frac{\pi }{2}}\ ,
\end{equation}
or that
\begin{equation}
\lim _{\lambda \rightarrow 0} k\ =\ p+{\frac{1}{2}}\ ,
\end{equation}
where $p$ is any non negative integer.  Using Eq. (2.6), it is easy to see that
one recovers the usual ($\ell = 0$) spectrum of the KG equation for $\alpha Z$
tending to 1/2 from below.  This means that the spectrum is continuous through
$\alpha Z = \frac{1}{2}$ for any value of $B \neq \frac{\pi}{2}$.
However, the slope of the energy curves (in $\lambda$ or $\alpha Z$) are not
continuous.  For small values of $\lambda$, one can show that the
eigenvalues correspond to the following behavior :
\begin{equation}
k\left({\lambda }\right)\simeq p+{\frac{1}{2}}+{\frac{\lambda }{B-\pi /2}
} \ .
\end{equation}
\par
On the other hand, for $B=\frac{\pi}{2}$, another state with no
correspondence with the spectrum
for $\alpha Z < \frac{1}{2}$ appears.  Its energy is given, from Eq. (2.15),
by the equation
\begin{equation}
\psi ({{\frac{1}{2}}-k})+\ln  2\sqrt
{1-{\frac{{E}^{2}}{{M}^{2}}}}-2\psi ({1})=0 \ ,
\end{equation}
and has the value $E/M \simeq  -0.049$ .\par
We now turn to the question of a possible instability of the system,
described by the KG
equation (2.1) with a pointlike attractive Coulomb interaction, with respect to
the spontaneous pair creation.  This would correspond to the existence of a
critical value of $Z=Z_c$ such that $E \left(Z_{c}\right) = -M$, or,
according to Eq. (2.6), $k = -\infty$.  To see whether such a solution
of Eq. (2.12)
exists for small values of $\lambda$, it is advantageous to transform
the third term of Eq. (2.12) by using the following formula
\cite{a} :
\begin{eqnarray}
{\rm Im}\ \ln \Gamma  \left(x+iy\right)\ &=&\ \arg \Gamma
 \left(x+iy\right)\ =\ \left(x - {\frac{1}{2}}\right)\ \arctan\
{\frac{y}{x}}\ +\ y\ \bigg\{\ \ln \ x\nonumber \\
+ {\frac{1}{2}} \ln  \left[1 +
\left({\frac{y}{x}}\right)^{2}\right] &-& y -
{\frac{y}{12\left({x}^{2}+{y}^{2}\right)}} + {\frac{1}{360}} b -
{\frac{1}{1260}} \left(b{a}_{1} + {ab}_{1}\right) - \cdots \
\bigg\} \ ,
\end{eqnarray}
with
\begin{eqnarray}
a \ &=&\ {\frac{x}{{x}^{2}\ +\ {y}^{2}}}\ a_1\ -\
{\frac{y}{{x}^{2}\ +\ {y}^{2}}}\ b_1\ ,\ \ \ b \ =\ {\frac{y}{{x}^{2}\ +\
{y}^{2}}}\ a_1\ +\ {\frac{x}{{x}^{2}\ +\ {y}^{2}}}\ b_1\ ,\nonumber \\
a_1\ &=&\ {\frac{{x}^{2}\ -\ {y}^{2}}
{{\left({{x}^{2}\ +\ {y}^{2}}\right)}^{2}}}\ ,\ \ \ b_1\
=\ {\frac{2xy}{{\left({{x}^{2}\ +\ {y}^{2}}\right)}^{2}}} \ .
\end{eqnarray}
One finds, using Eq. (2.12) for large values of $\vert k\vert$ and
Euler's constant $\gamma = -\psi(1)$ = 0.57721..., that the critical
value of $Z_c$ is given by
\begin{equation}
2{\lambda }_{c}\ \gamma \ +\ {\lambda }_{c} \ln
\left({2\alpha {Z}_{c}}\right)\ +\ {\cal O} \left({{\lambda
}_{c}^{2}}\right)\ =\ B\ -\ {\frac{\pi }{2}} ,
\end{equation}
where $\lambda _{c} =
\left(\alpha^{2}Z_{c}^{2} - 1/4\right)^{1/2}$.  Note the
cancellation between the two terms containing
$\ln \left(M^2 - E^2\right)$ in Eq. (2.12).  From Eq. (2.22), we see
that a critical value of $\alpha Z = \alpha Z_{c} > 1/2$
can only occur for $B \neq \frac{\pi}{2}$.
Also note that $\lambda_{c}$ is exactly $0$ if $B = \frac{\pi}{2}$, and will be
larger and larger if $B$ is increasing.  There is no solution of Eq. (2.22)
(at least for small values of $\lambda_c$) for $B \lesssim \frac{\pi}{2}$,
as the left-hand side should be positive.
The above comments are illustrated in Figs. 1 and 2
which show respectively how the energy of the lower bound states vary with
$\lambda$ for $B = 1.56 < \frac{\pi}{2}$ and $B = 1.58 >
\frac{\pi}{2}$ respectively.  One can see that no instability will occur
(i.e., no state with
$E = -M$) for $\lambda < 2$ in the first case $(B < \frac{\pi}{2})$,
while, for
$B >
\frac{\pi}{2}$, there is a state with energy $E = -M$ for $\lambda \simeq 0$.
Furthermore, the solution given by Eq. (2.19) corresponds (for small
$\lambda $) to the boundary case $B=\pi/2$ between the two domains
$B<\pi /2$ and $B>\pi /2$.\par
For the KG equation, the sign of the KG norm unambiguously
distinguishes between particle and possible antiparticle bound states.  We have
calculated numerically the norm of the states of $E \simeq -M$ wave
functions and established that they are indeed particle states.  We further
show in Fig. 3 the radial wave function corresponding to the lowest state for
$\alpha Z= 1$ and $B = 1.58$ (the dot-dashed curve in Fig. 2).  It, of course,
displays an infinite number of nodes consistent with the behaviour
(2.9), but the outermost of these nodes occurs at a distance representing a
rather small fraction (here of the order of one tenth) of the ``radius''
of the state.
Globally, the wave function has a ``nodeless'' structure, typical of a ground
state wavefunction.\par
We thus find that the stability property with respect to pair creation of
a Coulomb source with vanishing radius and charge
$Z >137/2$ crucially depends on the choice of the self-adjoint extension
characterized by the constant $B$. In principle, this constant should
be determined by comparison with experimental data. However,
the case of physical interest corresponds here to a finite radius $R$
of the source and  has been extensively discussed in the literature
\cite{prfkzp}. We
shall not pursue further the investigation of this academic problem.  Let us
simply note that the Coulomb field of a charge with radius $R$
satisfying $MR << 1$ is supercritical for $\alpha Z >
\alpha Z_{c} = 1/2$
\cite{bc}, so that the choice $B > \pi /2$ would be
mandatory to reproduce finite radii results.

\section{STRONGLY COUPLED POSITRONIUM}

We now consider the problem of two particles of equal mass $m$ and
opposite charges with spin
one-half, in mutual electromagnetic interaction.  Within RQCD the $^{1}S_{0}$
states of the system with the Todorov choice of the interaction
\cite{t} are described in the c.m. frame by the radial equation
\cite{bc,bcs} (see also the appendix) :
\begin{equation}
\left[{\ -\ {\frac{{d}^{2}}{{dr}^{2}}}\ +\ {m}_{ W}^{ 2}\ -\ {\left({
{\varepsilon }_{ W}\ - A(r) }\right) }^{2}\ }\right]\
 \varphi \ =\ 0 \ ,
\end{equation}
with
\begin{equation}
A(r) = - \frac{\alpha}{r}\ ,
\end{equation}
\begin{equation}
{\varepsilon }_{ W}\ =\ {\frac{{W}^{ 2} - {2m}^{ 2}}{2W}}\ ,
\end{equation}
\begin{equation}
{m}_{ W}\ =\ {\frac{{m}^{ 2}}{W}} \ ,
\end{equation}
$W$ being the c.m. energy of the two-body system.  Equation
(3.1) correctly describes the physical positronium $^1S_0$ energy
levels to order
$\alpha^{4}$ \cite{cvawb,sms}, with $\alpha$ being the fine structure
constant.  Our
interest is in the solution of Eq. (3.1) for large arbitrary values of
$\alpha$ (strongly coupled positronium).  From now on, we shall
consider $\alpha$ as a free parameter.\par
As Eq. (3.1) is very similar to the
Klein-Gordon equation (2.1), Case's method can be readily applied to it
to obtain its solutions in
a similar way. From Eq. (3.1) we find, for two square integrable
solutions ${\varphi }_{ 1}$ and ${\varphi }_{ 2}$ of energy $W_{ 1}$ and
$W_{ 2}$ ($W_1 \neq W_2$), the orthogonality condition :
\begin{equation}
\left({{\varphi }_{1},{\varphi }_{2}}\right)\ =\ {\frac{1}{2}} \left({{\varphi
}_{2}{'}{\varphi }_{1} - {\varphi }_{1}{'}{\varphi }_{2}}\right)
\bigg \vert _0^{\infty} \ =\ 0\ ,
\end{equation}
where the scalar product is now defined by :
\begin{equation}
\left( {\varphi }_{ 1} , {\varphi }_{ 2} \right)\ =\ \int_{0}^{\infty
} \left[\frac{{W}_{ 1} + {W}_{ 2}}{4} - A(r) \left( 1 +
\frac{{2m}^{ 2}}{{W}_{ 1} {W}_{ 2}} \right)\right]\
\varphi_{1}\varphi_{2}\ dr\ .
\end{equation}
The conserved norm $N$ corresponding to Eq. (3.6) is given by :
\begin{equation}
N\ =\ \int_{0}^{\infty }\left[{ {\frac{W}{2}} - A(r) \left({ 1 +
{\frac{{2m}^{ 2}}{{W}^{ 2}}} }\right) }\right]\ {\varphi }^{ 2}\ dr\ .
\end{equation}
\par
For $\alpha >1/2$, the solutions of Eq. (3.1), vanishing at $r=\infty$,
are given by :
\begin{equation}
\varphi  (r)\ =\ c\ {W}_{ \widetilde{k} ,\mu} \left( \rho \right)\  ,
\end{equation}
with
\begin{equation}
\widetilde{k}\ =\ \frac{\alpha  {\varepsilon }_{ W} }{\left( {m}_{W}^{2}
- {\varepsilon }_{W}^{2} \right)^{1/2}}\ =\
\frac {\alpha  (W^2 -2m^2)}{W(4m^2 - W^2)^{1/2}}\ ,
\end{equation}
\begin{equation}
\mu\ =\ i\ \lambda \  ,\ \ \ \lambda \ =\ \left( {\alpha }^ 2 -
{\frac{1}{4}} \right)^{1/2}\ ,
\end{equation}
\begin{equation}
\rho = 2 Kr\ ,
\end{equation}
\begin{equation}
K\ =\ \left( {m}_{W}^{2} - {\varepsilon }_{W}^{2} \right)^
{1/2}\ =\ \frac {1}{2}\left(4m^2 - W^2\right)^{1/2}\ .
\end{equation}
\par
As $r \rightarrow 0$, the wave function $\varphi (r)$ exhibits the
behavior described in Eqs. (2.8), (2.9) and (2.11) (with $k$ replaced by
$\widetilde k$). Choosing the same value of $B$ for all the wave functions
guarantees the validity of the orthogonality condition (3.5)
for $W_1 \neq W_2$. The corresponding energy spectrum is then given
by a relation similar to Eq. (2.12) :
\begin{equation}
\arg \Gamma \left({1-2i\lambda }\right)\ +\ {\frac{\pi }{2}}\ -\
\arg \Gamma  \left(\frac{1}{2}-i\lambda -\widetilde{k}\right)\ +\
\lambda  \ln \left({\frac{2K}{m}}\right)\ =\ B+n\pi \ \ .
\end{equation}
\par
It is easy to see, using Eqs. (2.13) and
(2.14), that, for small values of $\lambda$, there are solutions of Eq.
(3.13) corresponding to
\begin{equation}
\lim _{\lambda \rightarrow 0} \widetilde{k}\ =\ p + {\frac{1}{2}}\ ,
\end{equation}
where $p$ is any non negative integer, meaning that the $\alpha < \frac{1}{2}$
spectrum \cite{bc} is extended continuously through $\alpha = \frac{1}{2}$.\par
On the other hand, Eq. (3.13) possesses a new set of solutions with
energies located in the positive vicinity of zero. To exhibit them, we
consider small values of $\lambda$ and large negative values of
$\widetilde k$, corresponding to small and positive values of $W$
(see Eqs. (3.9) and (3.12)).
Using Eqs. (2.20) and (2.21), we can rewrite Eq. (3.13) in this case as
\begin{equation}
-2\lambda \ +\ \lambda  \ln (\frac{1}{2} -
\widetilde {k})\ +\ \lambda  \ln \left({\frac{2K}{m}}\right)\ +\ {\cal O}
\left({\lambda }^{2}\right)\ =\ B\ -\ {\frac{\pi }{2}}\ +\ n\pi \ ,
\end{equation}
or, using Eqs. (3.9) and (3.12), as
\begin{equation}
-2\lambda \ -\ \lambda  \ln \left({{\frac{W}{m}}\ \alpha }\right)\
+\ \lambda \ \ln 2\ +\ {\cal O} \left({{\lambda }^{2}}\right)\ =\ B\
-\ {\frac{\pi }{2}}\ +\ n\pi \ \ .
\end{equation}
There are thus solutions of vanishing $W$ as $\lambda \rightarrow 0$ :
\begin{equation}
-\lambda  \ln \left( {\frac{W}{m}}\ \alpha \right)\ \rightarrow \ B\ -\
{\frac{\pi }{\rm 2}}\ +\ n\pi \ \ ,
\end{equation}
i.e.,
\begin{equation}
{\frac{W}{m}}\ \simeq \ {\alpha }^{-1}\ {e}^{{\displaystyle -(B +
n\pi - \pi /2)/\lambda }} \ ,
\end{equation}
where $n$ = 0, 1, 2,... for $B > \pi/2$ and $n$ = 1, 2, 3,... for $B \leq
\pi/2$ (the reason being that the right-hand side of Eq. (3.17)
should be positive for small values of $W$).
\par
The absence of these new types of solution in the
Klein-Gordon equation case is due essentially to the fact
that in Eq. (2.12) the term $\ln ((M^2-E^2)^{1/2}/M)$ cancels, in the
limit $E\rightarrow -M$, the contribution coming from the large negative
values of $k$; no such cancellation exists in the present case, for the
equivalent parameter $K$ [Eq. (3.12)] remains finite and different from
zero in the limit $W\rightarrow 0$.\par
Equation (3.18) shows that, for a fixed value of $B$ and sufficiently
small $\lambda $, there is an
infinite number of states concentrated in a narrow band of energy and
having the zero energy as an accumulation point. Clearly, the zero energy
state is the new ground state of the bound state spectrum.
The energy spectrum given by Eq. (3.13) is plotted in Figs. 4 and 5
and illustrates the above comments again for values of
$B < \pi/2$ (Fig. 4) and $B > \pi/2$ (Fig. 5).
Only a few states corresponding to Eq. (3.18) are shown.\par

\section{THE FATE OF TACHYONS}
To complete the analysis of the energy spectrum, we discuss in this
section the question of the status of tachyonic states. The presence of
such states in the spectrum of physical states would invalidate the
conclusions of Sec. III, since the zero energy state would no longer
represent the ground state of the spectrum, the latter then displaying rather
the characteristics of an unstable vacuum.\par
Tachyons are characterized by negative values of $W^2$, i.e., by
imaginary values of $W$. Inspection of Eqs. (3.9) and (3.12) shows that
in this case $\widetilde k$ becomes imaginary, while $K$ remains real.\par
Equation (3.1) still has the Whittaker functions $W_{\widetilde k ,\mu}$
as normalizable solutions. We define
\begin{equation}
\widetilde k\ =\ i\eta\ ,
\end{equation}
and examine the behavior of the wave function near the origin; it is given
by Eq. (2.8) :
\begin{equation}
W_{\widetilde k ,\mu}\ \simeq\ \frac {\Gamma (-2i\lambda )}
{\Gamma (\frac {1}{2} - i\lambda -i\eta)}\ \rho ^{\frac {1}{2}+i\lambda }
\ +\ \frac {\Gamma (2i\lambda )}{\Gamma (\frac{1}{2} +i\lambda -i\eta)}\
\rho^{\frac {1}{2}-i\lambda }\ .
\end{equation}
\par
In order to display the main qualitative features of this type of solution,
we consider, in the following, large values of $\vert \eta \vert$,while
keeping $\lambda $ small. Then, the functions $\Gamma (\frac {1}{2} \pm
i\lambda -i\eta)$ can be expanded in terms of $i\lambda$ :
\begin{equation}
\Gamma (\frac {1}{2} \pm i\lambda -i\eta)\ \simeq\ \Gamma (\frac {1}{2} -i\eta)
\ e^{{\displaystyle \pm(i\lambda \ln \vert \eta \vert + \epsilon (\eta)
\lambda \pi /2)}}\ ,
\end{equation}
[$\epsilon (\eta)$ is the sign of $\eta$,] leading in turn to the
following behavior of $W_{\widetilde k , \mu}$ :
\begin{eqnarray}
W_{\widetilde k ,\mu}(\rho)\ \simeq \ \frac {1}{\Gamma (\frac {1}{2} -
i\eta)}\ \bigg \{& & \Gamma (-2i\lambda)\ e^{{\displaystyle i\lambda \ln
\vert \eta \vert + \epsilon (\eta) \lambda \pi /2}}\ \rho^{\frac {1}{2}
+i\lambda} \nonumber \\
&+&\ \Gamma (2i\lambda )\ e^{{\displaystyle -i\lambda \ln
\vert \eta \vert - \epsilon (\eta) \lambda \pi /2}}\ \rho ^{\frac {1}{2}
-i\lambda }\ \bigg \}\ .
\end{eqnarray}
\par
The wave function $\varphi $ has the same formal behavior as in Eqs.
(2.9) or (2.11), with $\beta $ defined as
\begin{equation}
\beta \ =\ \arg \Gamma (-2i\lambda ) + \lambda \ln \vert \eta \vert -
i\epsilon (\eta) \lambda \frac {\pi}{2}\ .
\end{equation}
Notice that $\beta$, and hence $B$, is complex.\par
In order to study the orthogonality conditions for these states, we observe
that when tachyons exist, there is a doubling of states : for each
``eigenvalue'' $i\eta$, the value $-i\eta$ is also a solution. The rules of
constructing scalar products for such states have been studied in
the two papers of
Ref. \cite{ssrst}. For a state with complex ``eigenvalue'' $i\eta$
 one also considers the state with the complex conjugate ``eigenvalue''
$-i\eta$, called the associated vector. While the norm of a state with
complex eigenvalue is zero, its scalar product with its associated vector
is nonzero \cite{ssrst}. Therefore, the following rule should be adopted
for the choice of admissible states. Define admissible states as those
corresponding to a definite sign of $\eta$ in the ``eigenvalues'' $i\eta$.
Then, the corresponding adjoint states in the scalar product are the
associated vectors (with ``eigenvalues'' $-i\eta$); these appear there
complex conjugated.\par
The analysis of Case can then be repeated for the tachyonic states. Once
the same complex value of $B$ is chosen for the tachyonic solutions,
then these will satisfy among themselves the orthogonality conditions
(3.5) and might constitute admissible states. [Notice that associated
vectors cannot be considered as admissible states, for, according to
Eq. (4.5), they would not have the same value of $B$; they appear only
as the adjoints of admissible states.] However, the orthogonality
condition (3.5) fails when considered between a tachyonic state and a
``normal'' state (with $W^2>0$ and $W>0$), because the coefficient $B$
cannot be chosen the same for both types of solution (it must be complex
for the former and real for the latter). This is the reason why tachyons
are rejected from the Hilbert space of physical states.\par
Therefore, we end up with the conclusion that the physical spectrum is
free of tachyons and its ground state is the zero energy state.\par

\section{THE LIMIT $\alpha \rightarrow \frac {1}{2} + \epsilon$ AND
SPONTANEOUS CHIRAL SYMMETRY BREAKING}
The occurrence of a zero mass ground state, with the quantum numbers of a
pseudoscalar boson, in the bound state spectrum is suggestive of
spontaneous chiral symmetry breaking. However, the oscillatory nature of
the wave functions near the origin makes it difficult to define their
couplings to the axial vector current. Furthermore, the accumulation
of an infinite number of states around the zero mass state [Eq. (3.18)]
does not allow the disentanglement of the ground state from the rest of the
states of its neighborhood.\par
Within these circumstances, the limiting value $\alpha =\frac {1}{2} +
\epsilon$ ($\epsilon =+0$) plays a particular role for a physical
interpretation of the theory. In this limit, the oscillating behaviors
of the type (2.9) or (2.11) disappear from the wave functions and,
according to Eq. (3.18), all the low mass states shrink to a single state
with zero mass. A definite mass gap appears between the zero mass ground
state and the other massive states of the spectrum.\par
To study in more detail the properties of the corresponding
wave functions, it is preferable to reanalyze Eq. (3.1) for the particular
value $\alpha =\frac {1}{2}+\epsilon$. The normalizable solutions are
the Whittaker functions $W_{\widetilde k,0}$, which behave near the origin
as
\begin{equation}
W_{\widetilde k,0} (\rho)\ \simeq \ -\frac {(2\rho)^{1/2}}
{\Gamma(-\widetilde k + \frac {1}{2})}\ \left(\ \ln2\rho +
\psi(-\widetilde k +\frac {1}{2}) - 2\psi(1)\ \right)\ ,
\end{equation}
the various parameters and variables being already defined in Eqs.
(3.9)-(3.12),
while the general behavior of the solutions of Eq. (3.1) is :
\begin{equation}
\varphi\ \sim r^{1/2} \left(a\ln(mr) + b\right)\ ,
\end{equation}
where $a$ and $b$ are constants.\par
The orthogonality condition (3.5) requires from the admissible
normalizable solutions to satisfy the condition
\begin{equation}
\frac {b}{a}\ =\ A\ ,
\end{equation}
$A$ being the same (arbitrary) constant for all the solutions.
Eq. (5.1) then yields the eigenvalue equation :
\begin{equation}
\ln(\frac {2K}{m})\ +\ \psi (-\widetilde k + \frac {1}{2})\ -\
2\psi (1)\ =\ A\ .
\end{equation}
\par
In order that the corresponding solutions be the limits of those found
for $\alpha > 1/2$ [Eqs. (3.14) and (3.18)], it is necessary that the
constant $A$ equal $+\infty$. One therefore finds again the solutions
(3.14), as well as the additional single solution $W=0$, corresponding to
$\widetilde k \rightarrow -\infty $.\par
The infinite value of $A$ means that for the solutions of the type
(3.14) the logarithmic piece in Eq. (5.2) is absent and the corresponding
wave functions behave as $r^{1/2}$ near the origin. This is not true for
the solution corresponding to $W=0$, for the limit $\widetilde k
\rightarrow -\infty$ cannot be straightforwardly taken in the Whittaker
function $W_{\widetilde k,0}$. To analyze more accurately the properties
of the corresponding
wave function, we consider small values of $W$ ($W =W_0 \simeq 0$),
make in Eq. (3.1) the change of variable
\begin{equation}
y\ =\ \frac{m^2 r}{W_0}\ ,
\end{equation}
and keep only leading terms, it being understood that the limit $W_0
\rightarrow 0$ should be taken at the end of calculations of physiacal
quantities.
Equation (3.1) becomes at leading order in $W_0$ :
\begin{equation}
\frac {d^2}{dy^2} \varphi\ -\ \frac {1}{y} \varphi \ +\ \frac {1}{4y^2}
\varphi\ =\ 0\ ,
\end{equation}
the solution of which is :
\begin{equation}
\varphi_0\ =\ c_0 m y^{\frac {1}{2}}\ e^{{\displaystyle -2y^{1/2}}}\
\Psi (\frac {1}{2},1,4y^{1/2})\ ,
\end{equation}
where $c_0$ is a dimensionless normalization constant and
$\Psi$ (also denoted by $U$ in the literature) is the confluent
hypergeometric function which behaves at infinity with a power law
\cite{as1}; its dominant behavior near the origin is given by
\begin{equation}
\lim _{x \rightarrow 0} \Psi(\frac {1}{2},1,x)\
=\ -\frac {1}{\Gamma (\frac{1}{2})}\ \left(\ \ln x + \psi (\frac {1}{2})
-2\psi (1)\ \right )\ .
\end{equation}
\par
The solution (5.7) is a nodeless function and represents the ground state
of the spectrum. [The function $\Psi (a,c;x)$ does not have positive zeros
for $a$ and $c$ real and either $a>0$ or $a-c+1>0$. Also, it can be
checked that the formal tachyonic solutions, found in Sec. IV, are
absent in the present case.]
It has a distribution-like behavior, due to the fact that
it is defined by the limiting procedure $W_0\rightarrow 0$ :
taking the limit $W_0 \rightarrow 0$, while keeping
$r$ fixed, shows that the wave function is actually peaked at the origin
(recall that the complete wave function is $\varphi /r$).\par
To compute the normalization constant of this wave function (as well as of
the others), it is necessary to reconstruct the whole sixteen-component
spinor wave function and to use its relationship with the
Bethe$-$Salpeter wave function, which ultimately fixes the normalization
coefficients. Some details of these calculations can be found in the
appendix . One finds for the norm $N$ of Eq. (3.17) the expression :
\begin{equation}
N\ =\ \frac {W^2}{8\pi }\ ,
\end{equation}
where $W$ is the mass of the bound state.\par
As to the massless state, after using the change of variable (5.5) in Eq.
(3.7), we find that the dominant contribution for small $W$ comes from the
third term in the integral. This leads to the following behavior of the
normalization constant $c_0$ of the wave function $\varphi _0 (y)$
[Eq. (5.7)] for small $W$ :
\begin{equation}
c_0\ \sim\ W_0^2/m^2\ .
\end{equation}
\par
We now turn to the calculation of the coupling constants of the bound
states to the axial vector current. These are defined as
\begin{equation}
<0|j_{\mu 5}^R(0)|P>\ =\ P_{\mu} F\ ,
\end{equation}
where $P$ is the four-momentum of the pseudoscalar state and
$j_{\mu 5}^R$ is the renormalized axial vector current. In general, in the
absence of anomalies, the axial vector current undergoes only a finite
multiplicative renormalization by radiative corrections \cite{pw}. This
feature is, however, the result of compensating contributions
from propagator and vertex renormalizations (with factors $Z_2$ and
$Z_A^{-1}$, respectively). More explicitly, one has
\begin{equation}
j_{\mu 5}\ =\ Z_2 Z_A^{-1} j_{\mu 5}^R\ ,
\end{equation}
where $j_{\mu 5}$ is the unrenormalized current, with $Z_2 Z_A^{-1}$
finite.\par
{}From operator product expansion \cite{w} and renormalization group
analysis \cite{pg} one finds:
\begin{equation}
<0|j_{05}^R|P>\ =\ -Z_A Tr \gamma _0 \gamma _5 \phi _{BS} (x)\bigg
\vert _{x\rightarrow 0} \ ,
\end{equation}
where $\phi _{BS}$ is the Bethe$-$Salpeter wave function.
Using the relationship of the constraint theory wave function $\psi$
with the Bethe$-$Salpeter wave function $\phi_{BS}$ [Eqs. (A6)-(A8)] one
obtains (in the c.m. frame) :
\begin{eqnarray}
<0|j_{05}^R|P> &=& -Z_A Tr \gamma _0 \gamma _5 \left(1+\frac {2\alpha }
{Wr}\right)^{-1/2}\ \psi (r) \bigg \vert _{r \rightarrow 0}\ ,\\
 & & \nonumber \\
WF&=& 2Z_A \left(1+\frac {2\alpha}{Wr}\right)^{-1/2}\ Tr \psi_4(r)\bigg
\vert _{r\rightarrow 0} \nonumber \\
&=& Z_A \frac {8m}{Wr}\varphi(r)\bigg \vert _{r \rightarrow 0}\ .
\end{eqnarray}
\par
The renormalization constant $Z_A$ should render finite the physical
coupling constants $F$.
Let $\Lambda $ be the ultra-violet cut-off (in momentum space) of the
four-dimensional theory and let $r_0$ be the corresponding short-distance
cut-off (in $x$-space) of the three-dimensional theory. We shall admit the
weak relation
\begin{equation}
\lim _{\Lambda \rightarrow \infty } r_0(\Lambda)\ =\ 0\ ,
\end{equation}
and shall transpose several known qualitative results of the four-dimensional
theory into the three-dimensional one. When $r_0 \neq 0$, the electron
has a bare mass $m_0(r_0)$ that vanishes with $r_0$ with some power $\nu $
\cite{bjw,adb} :
\begin{equation}
\lim _{r_0 \rightarrow 0} m_0(r_0) \ \sim \
m (mr_0)^{\nu }\ ,\ \ \ \ \nu > 0\ .
\end{equation}
For $m_0 \neq 0$, the Ward identities of the axial vector current imply
that the Goldstone boson acquires a mass $W_0$, which behaves in terms of
$m_0$ as \cite{p,h,njlgmorgl}
\begin{equation}
\lim _{r_0 \rightarrow 0} W_0^2 (r_0) \ \sim \ m_0(r_0) m\ .
\end{equation}
We shall assume that $\nu <2$; because of Eq. (5.18), it is
only in this case (including eventually the limiting case $\nu =2$) that
the wave function (5.7) can be consistently defined. As a matter of rough
comparison, the analog of $\nu $ in the four-dimensional theory, calculated
at the two loop level, is equal to $3\alpha/(2\pi ) + (3/4)(\alpha /(2\pi ))^2
\simeq 0.25$ for $\alpha = 1/2$ \cite{bjw,adb}.\par
We designate by $W_1$ and $F_1$ the mass and coupling constant of a massive
state of the bound state spectrum and by $W_0$ and $F_0$ the similar
quantities of the ground state (the Goldstone boson) when $r_0 \neq 0$.
Taking into account the behaviors of the corresponding wave functions near
the origin [$\varphi _1 \sim m c_1 (mr)^{1/2}$, $\varphi _0 \sim
mc_0 (m^2 r/W_0)^{1/2} \ln (m^2r_0/W_0)$], Eq. (5.15) leads for the two
cases to the following equations :
\begin{eqnarray}
F_1& =& Z_A \frac {8m^3c_1}{W_1^2 (mr_0)^{1/2}}\ ,\\
 & & \nonumber \\
F_0& =& Z_A \frac {8m^3 c_0 \ln (m^2r_0/W_0)}{W_0^2 (W_0 r_0)^{1/2}}\ .
\end{eqnarray}
In Eq. (5.19), $c_1$ is a nonvanishing normalization constant in the
limit $r_0 \rightarrow 0$, while the bahavior of $c_0$ in Eq. (5.20) is
given by Eq. (5.10).
In order to maintain finite the value of $F_0$ in the limit $r_0
\rightarrow 0$, we must have :
\begin{equation}
Z_A\ \sim \ (W_0r_0)^{1/2}/\ln(mr_0)^{-1}\ \sim \ (mr_0)^{\frac {1}{2}
+ \frac {\nu }{4}}/\ln(mr_0)^{-1}\ .
\end{equation}
Replacing $Z_A$ in Eq. (5.19) yields $F_1=0$. Therefore we
obtain, when $r_0\rightarrow 0$, the following behaviors of the coupling
constants :
\begin{equation}
F_0\ \neq 0\ ,\ \ \ \ F_1\ =\ 0\ .
\end{equation}
These are precisely the complementary conditions for having spontaneous
breakdown of chiral symmetry : only the Goldstone boson couples to the
axial vector current.\par
The behavior of $Z_A$ when $r_0 \rightarrow 0$, given by Eq. (5.21), is in
qualitative agreement with its behavior in the four-dimensional theory :
in the Feynman gauge, $Z_2$, and hence $Z_A$, vanishes
when the ultra-violet cut-off $\Lambda $ goes to infinity \cite{adb}.\par
We can also calculate the matrix elements of the divergence operator
$\partial ^{\mu} j_{\mu 5}$; actually, this should only lead to a check
of the covariance property of the formalism. We find :
\begin{eqnarray}
<0|\partial ^{\mu} j_{\mu 5}^R|P> &=& -iW^2 F \nonumber \\
&=& -Z_A 2im (1+\frac {2\alpha }{Wr})^{-1}\ Tr \gamma _5 \psi (r)
\bigg \vert _{r\rightarrow 0} \nonumber \\
&=& -Z_A 4im (1+\frac {2\alpha }{Wr})^{-1}\ Tr \psi _3(r) \bigg \vert
_{r\rightarrow 0} \nonumber \\
&=& - Z_A 8im \frac {\varphi }{r} \bigg \vert _{r\rightarrow 0}\ ,
\end{eqnarray}
which yields back Eq. (5.15). [In obtaining the above results, the wave
equations of $\psi $ have been used; similar calculations can be found in
the second paper of Ref. \cite{sms}.] According to Eqs. (5.22), for all
states of the bound state spectrum, the matrix elements (5.23) vanish,
thus ensuring axial vector current conservation.\par
We end up with the conclusion that, in the limit $\alpha =\frac {1}{2} +
\epsilon$, the theory displays the features of spontaneous chiral
symmetry breaking : i) the presence of a zero mass ground state in the
spectrum, with a mass gap with the rest of the bound states; ii) a
nonvanishing coupling of the Goldstone boson to the axial vector current
with a decoupling of the massive states from the latter.\par
For completeness, let us also describe the situation that results
from the limit $\alpha = \frac {1}{2} - \epsilon$. For $\alpha <
\frac {1}{2}$, the positronium spectrum has the usual structure
\cite{bc}, the same as for small $\alpha $, without massless bound
states, and the limit $\alpha \rightarrow \frac {1}{2}$ from below
does not introduce any qualitative changes, the ground state
remaining massive. In this case the renormalization constant $Z_A$ is
determined from the finiteness of $F_1$ [Eq. (5.19)]. The latter
should be different from zero; otherwise, the axial vector current
would be conserved and, in the absence of a Goldstone boson, chiral
symmetry would be realized through its normal mode, implying a parity
doubling of degenerate states; this is not realized in the bound
state spectrum; furthermore, in QED, with only electrons as massive
fermions, this also is not possible. We therefore conclude that for
$\alpha = \frac {1}{2} - \epsilon$, the axial vector current is not
conserved and chiral symmetry is explicitly broken by the electron
mass.\par

\section{SUMMARY AND DISCUSSION}
We applied Case's method of self-adjoint extension of singular potentials,
to the study of strongly coupled positronium in its pseudoscalar sector
in the framework of relativistic quantum constraint dynamics.
We found that, as the coupling constant $\alpha $ increases, the bound state
spectrum undergoes, at the critical value $\alpha =\alpha _c =1/2$, an
abrupt qualitative change. For $\alpha >\alpha _c$, the mass spectrum
displays, in addition to the existing states for $\alpha <\alpha _c$,
a new set of an infinite number of bound states, concentrated
in a narrow band starting at mass $W=0$. The bound states have
indefinitely oscillating wave functions near the origin.\par
In the limit $\alpha \rightarrow \alpha _c$ from above, the short-distance
oscillations disappear and the states accumulated around the zero mass
state, shrink to a single massless state, representing the ground
state of the spectrum, with a definite mass gap with the rest of the
states. This state has the required properties to represent a Goldstone
boson and hence it signals a transition to a new phase where chiral symmetry
is spontaneously broken.
This is also an expected possibility from the existence of an ultra-violet
stable fixed point in QED and is therefore suggestive of an identification
of the critical value $\alpha _c$ with the Gell$-$Mann$-$Low eigenvalue
$\alpha _0$. The fact that the two boundary values $\alpha _c - \epsilon$
and $\alpha _c +\epsilon$ correspond to different phases, the former
governing a phase where chiral symmetry is broken by the electron mass
term and the latter governing a phase where chiral symmetry is
spontaneously broken, necessitates the introduction of a similar distinction
for $\alpha _0$, with boundary values $\alpha _0 -\epsilon$ and $\alpha _0
+\epsilon$, with the search for the relevant domains in the theory.\par
In the model potential we were considering, the contribution of the
one-photon exchange diagram, besides kinematic factors, is represented
in the three-dimensional theory by $\alpha /r$, which means that no
distinction was made between the large- and short-distance behaviors
of the effective charge (the physical coupling constant $\alpha $ being,
in general, determined from the large distance behavior of the photon
propagator).
In order to be able to determine the domains of each of the above phases
with respect to the values of the physical coupling constant, we
define the effective charge in four-dimensional
momentum space \cite{gml,bjw,ad} as $\alpha _{eff} (-q^2/m^2) \equiv \alpha
d(-q^2/m^2,\alpha)$; $d$ is the Lorentz invariant part of the transverse
part of the photon propagator multiplied by $q^2$; $\alpha $ is the physical
coupling constant, measured at large distances : $\alpha _{eff}(0) =
\alpha $. The asymptotic value of $\alpha _{eff}$ is $\alpha _0$
[$\alpha _{eff} (\infty) =\alpha _0$], at which value the Gell$-$Mann$-$Low
function $\psi $ vanishes. Because of the positivity of the photon
two-point spectral function, one has in general the inequality
$\alpha _{eff} (-q^2/m^2)< \alpha _0$ for $q^2 <0$ \cite{ad},
which implies in
particular that $\alpha < \alpha _0$. \par
Another particular value of $\alpha $ is provided by the zero of
the Callan-Symanzik function $\beta $, which we denote by $\alpha _1$, with
$\alpha _1 < \alpha _0$, satisfying $\alpha _1 d^{as}(1,\alpha _1) =
\alpha _0$, where $d^{as}$ is the asymptotic part of $d$ \cite{ad}.
If radiative corrections are estimated to be of the order of $\alpha /\pi $
in general, then, for $\alpha \sim \alpha _0 =1/2$, they are of the order of
$20\%$ and $\alpha _1$ should be of the order of $0.4$.
It was shown by Adler \cite{ad} that, according to the ways of summing
diagrams, either $\alpha _1$ or $\alpha _0$ are essential singularities
for the corresponding defining functions ($\beta $ or $\psi $). The
value $\alpha _1$ appears then as a natural separation point between two
subdomains in the domain of variation $0<\alpha <\alpha _0$.
The following
scheme might provide a possible description of the conditions of
occurrences of each of the phases mentioned above.\par
When $0<\alpha <\alpha _1$, the asymptotic behavior of the photon
propagator is governed by $\alpha _0 -\epsilon$ and we are in the phase
where chiral symmetry is broken by the electron mass. In this phase, the
renormalized vertex function $m\overline \Gamma ^5$ (defined in Ref. \cite{p}),
corresponding to the divergence of the axial vector current,
is different from zero and the axial vector current is not conserved (in
the absence of anomalies). The structure of the positronium spectrum
is the same as for small values of $\alpha $, with a massive ground state.\par
When $\alpha $ jumps from $\alpha _1 -\epsilon$ to $\alpha _1 +\epsilon$,
this induces, through the relationship between the $\beta $ and $\psi $
functions, a similar jump of the bare coupling constant
from $\alpha _0 -\epsilon$ to $\alpha _0 +\epsilon$ and we enter in the
phase where chiral symmetry is spontaneously broken. In this phase
$m\overline \Gamma ^5$ is identically zero and the axial vector current
is conserved. The positronium spectrum has now, in addition to the existing
states for $\alpha <\alpha _1$, a massless ground state.\par
When $\alpha > \alpha _0$, because of the positivity condition already
mentioned, the unitarity of the theory breaks dowm.\par
Our conclusions also join those obtained by Miransky {\it et al.}
\cite{fgmmmf} from the Bethe$-$Salpeter equation, who conjectured that
the critical value $\alpha _c$ should be identified with the
Gell$-$Mann$-$Low eigenvalue. The difference in the numerical values
of $\alpha _c$ found in the two approaches ($\alpha _c =1/2$ here and
$\alpha _c =\pi/4$ in the ladder approximation of the Bethe$-$Salpeter
equation in the Feynman gauge) is presumably related to the different
approximations used in the kernels of the bound state wave equations.
The Todorov potential, used in the present approach, takes into account
multi-photon exchange diagrams and correctly reproduces the physical
positronium and muonium spectra to order $\alpha ^4$ \cite{cvawb,sms};
this is not the case for the ladder approximation of the Bethe$-$Salpeter
equation in covariant gauges.\par
In the course of the present analysis, the effects of anomalies were
ignored; these are known to modify the Ward identities of the axial
vector current \cite{ad2}. However, in QED, it turns out that these
effects disappear at zero momentum transfer \cite{adb2} and hence they do
not seem to be able to give a mass to the Goldstone boson, when the
latter exists. It is only in non-abelian gauge theories that
nonperturbative effects, like those of instantons, succeed,
through the anomalous Ward identities, in providing the Goldstone
boson with a mass \cite{th}.\par
Finally, a comment on the structure of the Goldstone boson is in order.
This state, in the present mechanism of chiral symmetry breaking, is not
of the same nature as that of the massive states. In particular, it does
not result from the continuous decrease of the mass of a massive state
down to zero, when the physical coupling constant increases,
but rather appears abruptly as a new type of solution
when the physical coupling constant exceeds a critical value.\par

\acknowledgments The work of M.B. was supported by the National Fund for
Scientific Research, Belgium. \par

\appendix
\section*{}
The wave equations of constraint theory for a fermion-antifermion system
can be written in the form \cite{sms}
\begin{mathletters}
\begin{eqnarray}
(\gamma _1.p_1 - m_1)\ \widetilde \Psi &=& (-\gamma _2.p_2 + m_2)
\ \widetilde V\ \widetilde \Psi\ ,\\
(-\gamma _2.p_2 - m_2)\ \widetilde \Psi &=& (\gamma_ 1.p_1 + m_1)
\ \widetilde V\ \widetilde \Psi\ ,
\end{eqnarray}
\end{mathletters}
where $\widetilde \Psi$ is a sixteen-component spinor wave function of rank
two and is represented as a $4\times 4$ matrix; the Dirac matrices
$\gamma _2$ act on $\widetilde \Psi$ from the right. The compatibility
condition of the two equations (A1) allow one to eliminate the relative
time variable and to define an internal three-dimensional wave function.\par
After using the parametrization
\begin{equation}
\widetilde V\ =\ \tanh V
\end{equation}
and making the change of function
\begin{equation}
\widetilde \Psi \ =\ (\cosh V)\ \Psi \ ,
\end{equation}
the norm of the internal three-dimensional wave function, denoted by
$\psi $, becomes (in the c.m. frame) :
\begin{equation}
\int d^3 {\bf x}\ Tr \left \{ \psi ^{\dag} [1+4\gamma _{10}
\gamma _{20} P_0^2 \frac {\partial V}{\partial P^2} ]\psi \right \}\
=\ 2P_0\ ,
\end{equation}
where $P_{\mu}$ is the total four-momentum of the system.\par
In this representation the Todorov potential \cite{t} takes the form
(in the Feynman gauge):
\begin{equation}
V\ =\ \gamma _1.\gamma _2 \frac {1}{4} \ln \left (1+\frac {2\alpha}
{Wr} \right )\ ,\ \ \ \ W = \sqrt {P^2}\ .
\end{equation}
\par
Equations (A1) can be solved by first decomposing $\psi $ (the
internal part of $\Psi $) on the basis of
the matrices $1,\ \gamma _0,\ \gamma _5$ and $\gamma _0 \gamma _5$ :
\begin{equation}
\psi \ =\ \psi _1 \ +\ \gamma _0\psi _2\ +\ \gamma _5\psi _3\ +
\ \gamma _0\gamma _5\psi _4\ ,
\end{equation}
with $\psi _i$ ($i=1,\ldots ,4$) considered as $2\times 2$ matrices in
the spin subspace. The relationships of these components with the wave
function $\varphi $ used throughout the text are (for the equal mass
case and the quantum numbers $s=0,\ \ell =0,\ j=0$) :
\begin{eqnarray}
\psi _1 &=& \frac {2}{W} ({\bf s}_1 - {\bf s}_2).{\bf p}\
\frac {\varphi }{r}\ ,
\ \ \ \ \ \ \psi _2 = 0\ ,\nonumber \\
\psi _3 &=& \left (1+\frac {2\alpha }{Wr}\right )\
\frac {\varphi }{r}\ ,\ \ \ \ \ \ \ \ \
\psi _4 = \frac {2m}{W} \left (1+\frac {2\alpha }{Wr}\right )^{1/2}\
\frac {\varphi }{r}\ ,
\end{eqnarray}
where ${\bf s}_1$ and ${\bf s}_2$ are the spin operators of particles 1
and 2 in the $2\times 2$ component subspace of the $\psi _i$'s.\par
In perturbation theory, a relationship can be established by means of an
iterative series, between Eqs. (A1) and the Bethe$-$Salpeter equation
\cite{s}.
In general, the potential $\widetilde V$ is a three-dimensional nonlocal
operator in $x$-space, but becomes a  local function when appropriate
approximations are used. In particular, when the nonlocal operator
$(m^2 + {\bf p}^2)^{-1/2}$ is replaced by a mean value like
$(m^2 + <{\bf p}^2>)^{-1/2}$, $\widetilde V$ becomes local in ${\bf x}$
(in the c.m. frame) and dependent on $(m^2 + <{\bf p}^2>)^{-1/2}$.\par
The Todorov potential (A5) results, however, from a slightly different
approximation : it is a function of ${\bf x}$ and $W/2$, rather than of
${\bf x}$ and $(m^2 + <{\bf p}^2>)^{-1/2}$. One should then replace the
latter quantity by $W/2$. It turns out that this approximation provides
even better results, since the Todorov potential reproduces the correct
spectrum to order $\alpha ^4$ for positronium and muonium \cite{cvawb,sms}.
In this approximation, at zero relative time ($x^0 = 0$), the relationship
between the Bethe$-$Salpeter wave function $\phi _{BS}$ and $\psi $
takes the form :
\begin{equation}
\phi _{BS} (x^0 =0,{\bf x})\ =\ e^{\displaystyle \gamma _{10} \gamma _{20}
V}\ \psi ({\bf x})\ ,
\end{equation}
where $V$ is given in Eq. (A5). The normalization constant in the
right-hand side of Eq. (A4) takes account of this relationship.\par

\begin{figure} \caption{$\ell = 0$ spectrum generated by Case's method for
solving the KG equation (2.1) in the strong coupling case $({\alpha Z
> 1/2})$, with Case's constant $B=1.56$.  The variable
$\lambda$ is defined by Eq. (2.7).}
\label{autonum}
\end{figure}

\begin{figure} \caption{Same as Fig. 1, with $B=1.58$.  The eigenvalues for all
states, except the lowest one, are continuously continuing the eigenvalues for
$\alpha Z <1/2$.}
\end{figure}

\begin{figure} \caption{Wave function of the lowest state generated by Case's
method for solving the KG equation (2.1), with $\alpha Z = 1$ and Case's
constant $B = 1.58$, corresponding to the dot-dashed curve in Fig. 2.}
\end{figure}

\begin{figure} \caption{$\ell = 0$ spectrum generated by Case's method for
solving the RQCD equation (3.1) in the strong coupling case $(\alpha
 >1/2)$ and for Case's constant $B=1.57$.  The $\lambda$
parameter is defined by Eq. (3.10).  All curves above and including the heavy
bold curve are connected continuously to the spectrum obtained for $\alpha$
tending to $1/2$ from below.  As indicated in the text, all the other
curves are merging to the origin of the axes as $\lambda \rightarrow 0$ (only
the explicitly calculated parts are shown).}
\end{figure}

\begin{figure} \caption{Similar results as in Fig. 4, but for $B = 1.58$.
The same conventions have been used. }
\end{figure}

\end{document}